\documentclass[sn-standardnature]{sn-jnl}

\jyear{2022}%

\theoremstyle{thmstyleone}%
\theoremstyle{thmstyletwo}%

\theoremstyle{thmstylethree}%

\usepackage{lineno}

\begin{document}

\title[ ]{Nonreciprocal Charge Transport in Topological Kagome Superconductor CsV\textsubscript{3}Sb\textsubscript{5}}


\author*[1,2]{\fnm{Yueshen} \sur{Wu}}\email{wuysh@shanghaitech.edu.cn.}
\equalcont{These authors contributed equally to this work.}

\author[1,2]{\fnm{Qi} \sur{Wang}}
\equalcont{These authors contributed equally to this work.}

\author[1,2]{\fnm{Xiang} \sur{Zhou}}

\author[1,2]{\fnm{Jinghui} \sur{Wang}}

\author[1,2]{\fnm{Peng} \sur{Dong}}

\author[1,2]{\fnm{Jiadian} \sur{He}}

\author[1,2]{\fnm{Yifan} \sur{Ding}}

\author[1,2]{\fnm{Bolun} \sur{Teng}}

\author[1,2]{\fnm{Yiwen} \sur{Zhang}}

\author[1,2]{\fnm{Yifei} \sur{Li}}

\author[1]{\fnm{Chenglong} \sur{Zhao}}

\author[1]{\fnm{Hongti} \sur{Zhang}}

\author[1,2]{\fnm{Jianpeng} \sur{Liu}}

\author*[1,2,3]{\fnm{Yanpeng} \sur{Qi}}\email{qiyp@shanghaitech.edu.cn}

\author[4]{\fnm{Kenji} \sur{Watanabe}}

\author[5]{\fnm{Takashi} \sur{Taniguchi}}

\author*[1,2,6,7]{\fnm{Jun} \sur{Li}}\email{lijun3@shanghaitech.edu.cn}

\affil*[1]{\orgdiv{School of Physical Science and Technology}, \orgname{ShanghaiTech University}, \orgaddress{\city{Shanghai}, \postcode{201210}, \country{China}}}

\affil*[2]{\orgdiv{ShanghaiTech Laboratory for Topological Physics}, \orgname{ShanghaiTech University}, \orgaddress{\city{Shanghai}, \postcode{201210}, \country{China}}}

\affil[3]{\orgdiv{Shanghai Key Laboratory of High-resolution Electron Microscopy}, \orgname{ShanghaiTech University}, \orgaddress{\city{Shanghai}, \postcode{201210}, \country{China}}}

\affil[4]{\orgdiv{Research Center for Functional Materials}, \orgname{National Institute for Materials Science}, \orgaddress{\city{Tsukuba}, \postcode{305-0044}, \country{Japan}}}

\affil[5]{\orgdiv{International Center for Materials Nanoarchitectonics}, \orgname{National Institute for Materials Science}, \orgaddress{\city{Tsukuba}, \postcode{305-0044}, \country{Japan}}}

\affil[6]{\orgdiv{State Key Laboratory of Functional Materials for Informatics}, \orgname{Shanghai Institute of Microsystem and Information Technology, Chinese Academy of Sciences}, \orgaddress{\street{865 Changning Road}, \city{Shanghai}, \postcode{200050}, \country{China}}}

\affil[7]{\orgdiv{Wuhan National High Magnetic Field Center}, \orgname{Huazhong University of Science \& Technology}, \orgaddress{\city{Wuhan}, \postcode{430074}, \country{China}}}

\abstract{Nonreciprocal charge transport phenomena are widely studied in two-dimensional superconductors, which demonstrate unidirectional-anisotropy magnetoresistances as a result of symmetry breaking. Here, we report a strong nonreciprocal transport phenomenon in superconducting CsV\textsubscript{3}Sb\textsubscript{5} thin flakes. The second harmonic voltages, mainly originating from the rectification effect of vortex motion, are unambiguously developed with in-plane and out-of-plane magnetic fields, and their magnitudes are comparable to those in noncentrosymmetric superconductors. The second harmonic magnetoresistances split into several peaks and some of them reverse their signs by ramping the magnetic field or the current within the superconducting transition. The nonreciprocity suggests a strong asymmetry in CsV\textsubscript{3}Sb\textsubscript{5}. The centrosymmetric structure and symmetric electronic phases in CsV\textsubscript{3}Sb\textsubscript{5} can hardly induce the distinct nonreciprocal transport phenomenon, which could be correlated to a symmetry breaking from an unconventional superconducting order parameter symmetry.}

\maketitle

\section*{Introduction}

The nonreciprocal charge transport phenomenon, also called magnetochiral anisotropy \cite{rikken1997observation} or unidirectional magnetoresistance \cite{zhou2021sign}, results from inversion and time-reversal symmetry breaking that can modulate the energy dispersion with unequal electron energies of opposite momentum \textbf{k} and $-$\textbf{k} or spin $s$ and $-s$ \cite{Tokura}. Based on this phenomenon, a wide range of potential applications such as rectifiers, alternating-direct-current converters, and photodetectors could be engineered \cite{Villegas1188,moshchalkov2010nanoscience,MCAZhang,nagulu2020non}. In superconductors, the nonreciprocal signal can be considerably enhanced in the superconducting state by several orders of magnitude due to the significant difference between Fermi energy and superconducting energy gaps \cite{Liueabc7628,PhysRevB.98.054510,daido2022intrinsic}, for which the nonreciprocal transport is related to the chirality of supercurrent or Cooper pairs. Up to now, nonreciprocal transport measurements have been performed in various superconductors such as the noncentrosymmetric MoS\textsubscript{2} \cite{MoS2PRR,Liueabc7628}, NbSe\textsubscript{2} \cite{MCAZhang}, Nb/V/Ta superlattice \cite{MCAAndo}, the polar superconductor SrTiO\textsubscript{3} \cite{Itahashieaay9120}, and the artificially engineered MoGe superconducting thin films \cite{lyu2021superconducting},all of which are basically related to the symmetry breaking in structures, electronic states, or even anisotropic friction forces on vortex flowing. On the other hand, the intrinsic vortex motion behavior correlated with the superconducting gap symmetry can also contribute to the nonreciprocity \cite{Liueabc7628}. In addition, such nonreciprocal transport phenomenon has been observed in the topological superconducting heterostructure of Bi\textsubscript{2}Te\textsubscript{3}/FeTe \cite{MCAYasuda}. Thus, these nonreciprocal responses can be studied as a probe to detect the chirality in an anisotropic superconducting states.

Recently, a series of kagome metals $A$V\textsubscript{3}Sb\textsubscript{5}($A$ = K, Rb, Cs) \cite{PhysRevLett.125.247002} have attracted increasing attentions with the unique coexistence of topological states, charge density wave (CDW), and multi-band superconductivity. Among these crystals, CsV\textsubscript{3}Sb\textsubscript{5} reveals the highest of 2.5~K \cite{PhysRevLett.127.177001}. Up to now, fruitful experiments have been actualized to understand the superconducting gap symmetry of CsV\textsubscript{3}Sb\textsubscript{5},  including thermal conductivity \cite{zhao2021nodal}, scanning tunneling spectroscopy measurements \cite{xu2021multiband}, and penetration length measurements \cite{Nodeless}. However, these results provided contradictory conclusions on the gap symmetry  \cite{PhysRevLett.127.177001,ding2022diagnosis,kim2022monolayer,jiang2022electronic,Mu_2021_s_wave,zhao2021nodal,xu2021multiband,Nodeless,Shunli,RN985,RN986,chiral,PhysRevLett.127.046401,feng2021chiral,yu2021evidence}. On the other hand, since the superconductivity coexists with CDW, their interplay should be also taken into account. The investigation of these two competitive orders  is one of research focuses in CsV\textsubscript{3}Sb\textsubscript{5} superconductor \cite{kang2022microscopic,STMCDW,NMR,li2022conjoined,li2021CDW,PhysRevX.11.041030,PhysRevLett.128.036402}. For instance, the observation of chiral CDW provides a possible mechanism for unconventional superconductivity \cite{chiral}, which induces a pair density wave phase below the superconducting transition temperature \cite{PDW,ge2022discovery}. Nevertheless, by applying high pressure, an unusual observation of two superconducting domes was found inside and outside the CDW phase \cite{PhysRevLett.126.247001,PhysRevB.103.224513,Pressureyu,Chen_2021,wangqi}, indicating the competition between CDW and superconductivity. In addition, CsV\textsubscript{3}Sb\textsubscript{5} is also a $\bf{Z}$\textsubscript{2}
topological metal in which the Dirac line locates between $\bf{K}$ and $\bf{H}$ points, and the topological surface states were confirmed by the angle-resolved photon electron spectroscopy (ARPES) \cite{PhysRevLett.125.247002,HU2021}. Moreover, evidence of anomalous Hall effect has been observed in the CDW state, which was attributed to the symmetry breaking of the band structures \cite{wang2021electronic,PhysRevB.104.L041103,fu2021quantum}. To address these issues, one of the most promising paths is investigating the possible chirality of Cooper pairs and vortex dynamics, which has not yet been reported.

Here, we report the study of the nonreciprocal charge transport of mechanically exfoliated CsV\textsubscript{3}Sb\textsubscript{5} thin flakes. As reducing the thickness, the superconducting transition temperature ($T_{\textrm{c}}$) is firstly enhanced but then suppressed, competing with the emergence of CDW. The nonreciprocal charge transport is observed when the magnetic field is aligned both within the $ab$-plane and along the $c$-axis. The second harmonic magnetoresistance splits into several peaks and reverses their signs by ramping the magnetic field or the current, suggesting a tunable rectification effect. Our results reveal detailed information about the chirality and vortex dynamics.

\section*{Results}

\subsection*{Sample characterization}

\begin{figure*}[h]
\centering
\includegraphics[width=0.8\textwidth]{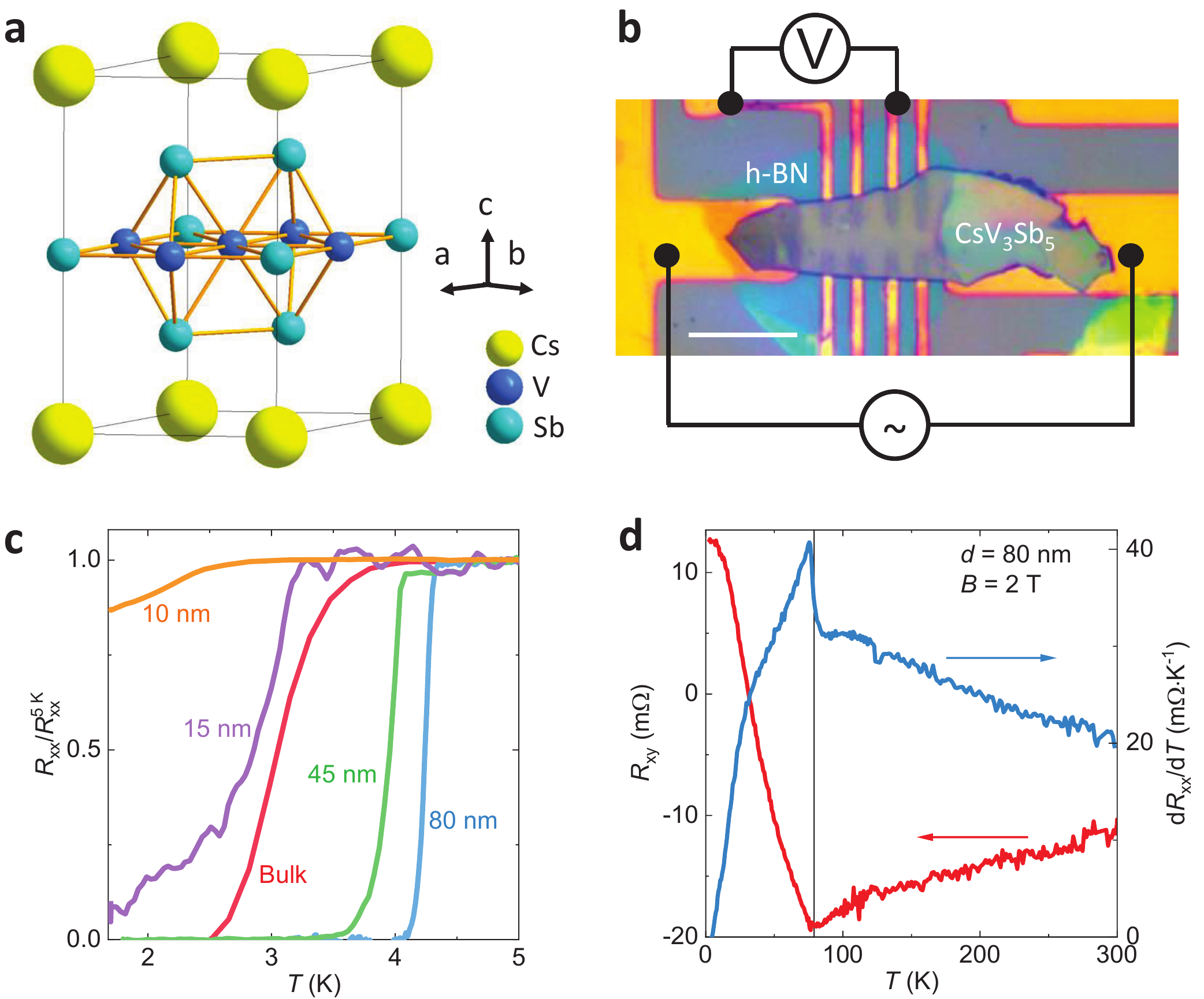}
\caption{\label{fig1} 
\textbf{Characterization of superconductivity. } \textbf{a} Crystal structure of CsV\protect\textsubscript{3}Sb\protect\textsubscript{5}. \textbf{b} Optical image of a sample covered with $h$-BN. Scale bar: 20 $\mathrm{\mu}$m. \textbf{c} Normalized resistance as a function of temperature with different thickness. \textbf{d} Temperature dependence of Hall resistance (red) and differential resistance (blue) of a 80-nm-thick sample under $H=$ 2 T. Here the resistance is measured with direct current.
}
\end{figure*}

CsV\textsubscript{3}Sb\textsubscript{5} is a layered material with a V-kagome lattice located in the middle of the unit cell shown in Fig. \ref{fig1}a. Thanks to the weakly bonding strength between Cs and Sb atoms, atomically thin crystals can be obtained by the mechanical exfoliation method. Thin flakes were covered by few-layer BN, and then transferred onto the pre-patterned gold leads as shown in Fig. \ref{fig1}b. Before the nonreciprocity measurements, we first examined the onset and zero resistance  $T_{\textrm{c}}$s, and their thickness dependence is shown in Fig. \ref{fig1}c. The onset $T_{\textrm{c}}$ of bulk crystal is 3.5 K, and the $T_{\textrm{c}}$ can be enhanced up to 4.25 K by shrinking the thickness of the crystal into 80~nm, followed by a decrease in $T_{\textrm{c}}$ to 2.5 K for the 10-nm-thick one. The $T_{\textrm{c}}$ of zero-resistance behaves the similarly behavior. The nonmonotonically evolution of $T_{\textrm{c}}$ with reducing layers may be correlated to the competition with CDW order. Particularly, in the 80-nm-thick sample, the transition temperature of CDW is reduced to 77 K, which can be confirmed by both Hall resistances and d$R$/d$T$ as shown in Fig. \ref{fig1}d. This phenomenon has been well discussed in recent reports \cite{song2021competing,song2021competition}, in which the surface oxidation that occurred during the sample fabrication process was atrributed to additional carrier doping and $T_{\textrm{c}}$ modification \cite{song2021competition}. In our present work, however, a piece of BN was covered onto the thin flakes to avoid oxidation, the $T_{\textrm{c}}$ is generally related to the thickness, we can hardly ascribe to the sample degradation. Here, we focus on the samples with $T_{\textrm{c}}$ of above 4 K (a 45-nm-thick device in the manuscript and others in Supplementary Figure 1 and 2) to study the vortex dynamics and the chirality of superconductivity from the nonreciprocal charge transport by probing the second harmonic response.

\subsection*{Nonreciprocal magnetoresistance}

\begin{figure*}[h]
\centering
\includegraphics[width=0.8\textwidth]{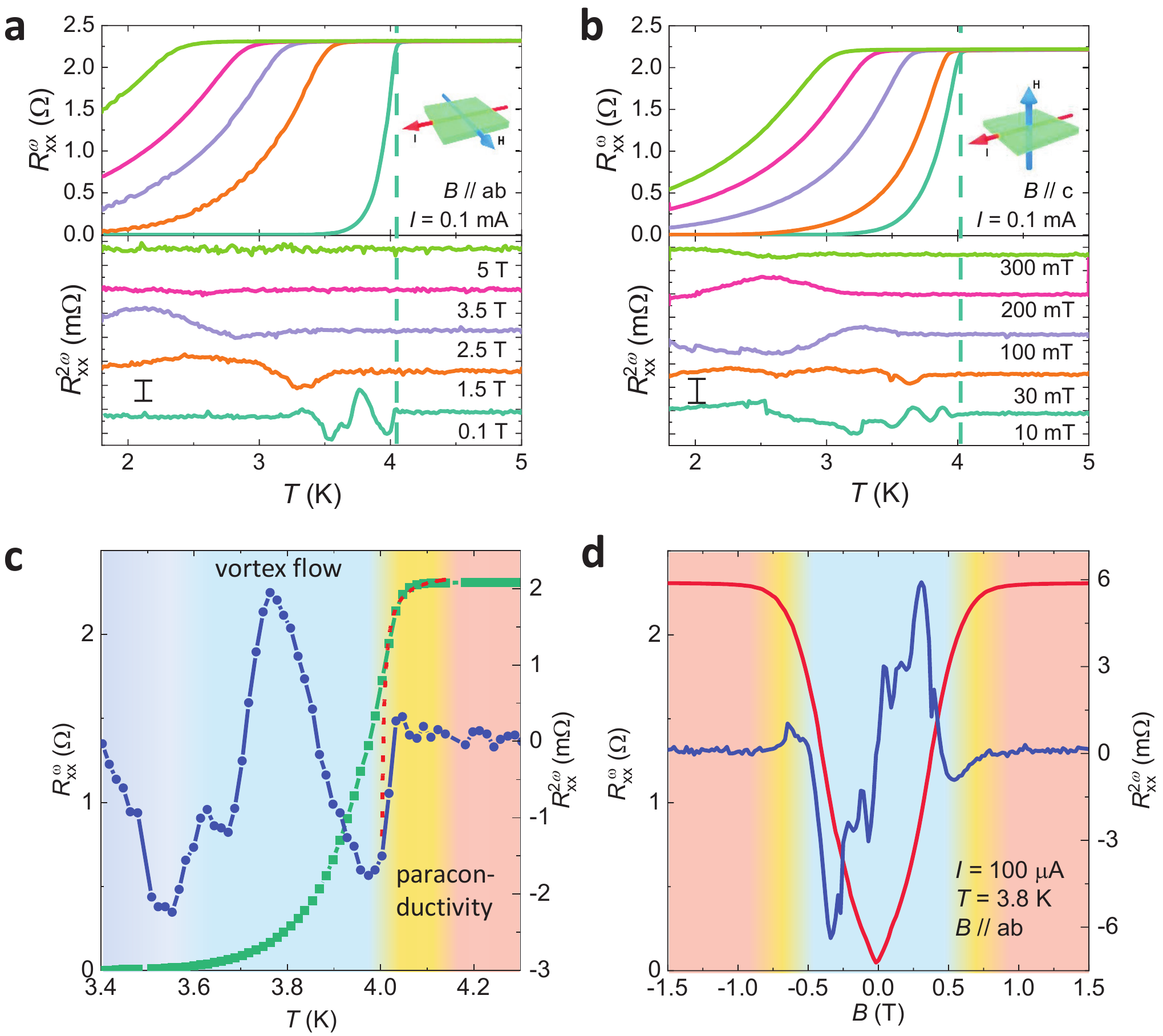}
\caption{
\label{fig2} 
\textbf{Anisotropic nonreciprocal magnetoresistance of a 45-nm-thick sample.} (a) (b) $R-T$ curves of first (top) and second (bottom) harmonic signal under $B$ applied along the $ab$-plane and the $c$-axis, respectively. Scale bar: 2 m$\Omega$.  The dashed line is the guideline of onset temperature of second harmonic signal. (c) The transition regions for $R_\textrm{xx}^{2\omega} -T$ curve under $B=0.1$ T. Red dashed line is the Aslamazov-Larkin fitting curve of paraconductivity contribution. (d) The $R_\textrm{xx}^{2\omega} -B$ and $R_\textrm{xx}^{\omega} -B$ curves at 3.8 K, where $B$ is applied within the $ab$-plane.
}
\end{figure*}

The first and second harmonic voltages ($V^{\omega}$ and $V^{2\omega}$) were detected using lock-in amplifier. According to the definition of nonreciprocity, the total voltage $V$ can be considered as $V=V^{\omega}+V^{2\omega}=V^{\omega}\left(1+\gamma BI\right)$, where, the coefficient $\gamma$ is equal to $2V^{2\omega}/V^{\omega}BI$.
Since the electronic structure of CsV\textsubscript{3}Sb\textsubscript{5} is quasi-two-dimensional, the nonreciprocal response could be anisotropic from in-plane to out-of-plane. We then measured and calculated the temperature dependence of the first and the second harmonic resistances ($R_\textrm{xx}^{\omega}=V^{\omega}/I_\textrm{0}$ and $R_\textrm{xx}^{2\omega}=V^{2\omega}/I_\textrm{0}$) simultaneously for both $B/\mkern-4mu/ab$-plane and $B/\mkern-4mu/c$-axis cases as shown in Fig. \ref{fig2}a and \ref{fig2}b. At low magnetic field limit for $B/\mkern-4mu/ab$-plane case, the nonreciprocal resistance immediately develops when $R_\textrm{xx}$ starts to decrease as marked in the green dashed line. On the other hand, the $R_\textrm{xx}^{2\omega}$ disappears at high magnetic field limit. Similar behavior is observed for $B/\mkern-4mu/c$-axis case as well. Another important feature is that several times of sign reversals appear at both $R_\textrm{xx}^{2\omega}$-$T$ and $R_\textrm{xx}^{2\omega}$-$B$ curves, which will be discussed in detail below. It should be noted that the nonreciprocal magnetoresistance is detected just within the superconductivity transition region. Thus, we focus on the investigation between nonreciprocal magnetoresistance and superconductivity.

Generally, the superconducting transition of the type-II superconductor consists of two regions, separated by a mean-field transition temperature $T_{\textrm{c}}^{\textrm{m}}$ \cite{Itahashieaay9120,MCAYasuda}. The $T_{\textrm{c}}^{\textrm{m}}$ can be determined as shown in Fig. \ref{fig2}c by using the Aslamazov-Larkin term $\sigma_\textrm{2D}=A(T-T_{\textrm{c}}^{\textrm{m}})^{-1}$, where $\sigma_\textrm{2D}$ is the two-dimensional conductivity and $A$ is a fitting parameter. Thus, we can confirm the $T_{\textrm{c}}^{\textrm{m}}$ is about 4.00 K. Above the $T_{\textrm{c}}^{\textrm{m}}$, the transport properties are dominated by the amplitude fluctuation of order parameters which causes a paraconductivity state. Below the $T_{\textrm{c}}^{\textrm{m}}$, a phase fluctuation generates vortices motion, which takes charge of the transport properties. It is worth noting that the resistance is still nonzero until about 3.45 K due to the current-driven vortex motion. Below 3.45 K, the vortices are frozen as the vortex solid state (or vortex lattice for the type-II superconductors) and the system is in superconductivity state completely. Meanwhile, the temperature-dependent $R_\textrm{xx}^{2\omega}$ curve demonstrates four characteristic regions as well, namely, normal state, paraconductivity, vortex motion, and vortex solid state according to those of the $R_\textrm{xx}^{\omega}-T$ curve. For the magnetoresistance measurements, the $R_\textrm{xx}^{\omega}$ reveals a typical magnetic-field-dependent behavior as shown in Fig. \ref{fig2}d, while the $R_\textrm{xx}^{2\omega}$ curve demonstrates antisymmetric to the magnetic field, namely, a typical characteristic of nonreciprocity.

\begin{figure}[h]
\centering
\includegraphics[width=0.8\columnwidth]{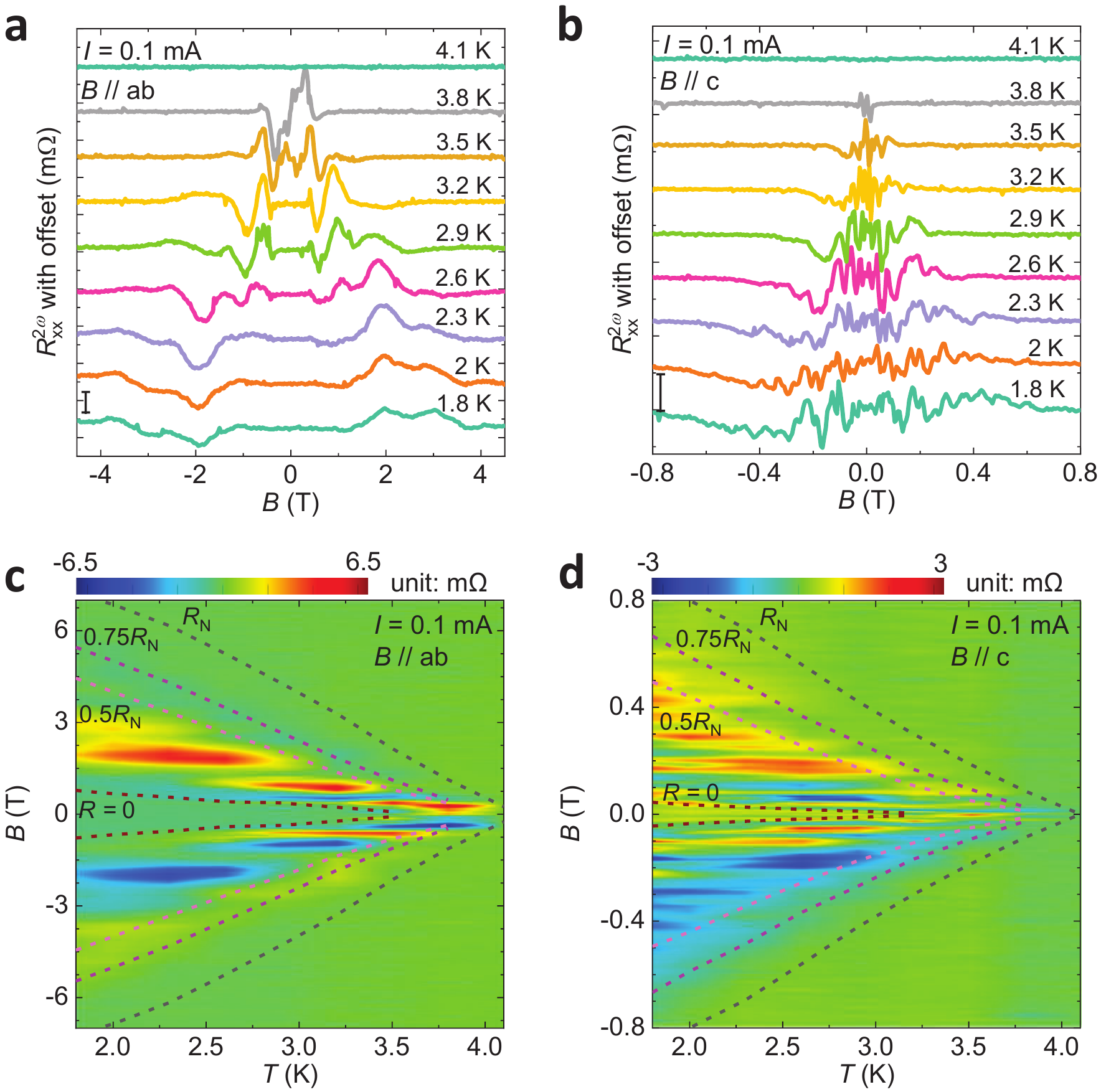}
\caption{
\label{fig3} 
\textbf{The magnetic field and temperature dependence of $R_\textrm{xx}^{2\omega}$ of the 45-nm-thick sample.} \textbf{a} and \textbf{b} are the second harmonic magnetoresistance with the magnetic field applied along the $ab$-plane and the $c$-axis at different temperatures. Scale bar: 2.5 m$\Omega$. \textbf{c} and \textbf{d} are the color mapping plot of \textbf{a} and \textbf{b}, respectively. The dashed lines are the contours of first harmonic resistance $R^{\omega}$ of zero resistance, half of the normal state resistance (0.5$R_\textrm{N}$), 0.75$R_\textrm{N}$ and onset of transition.
}
\end{figure}

To explore the magnetochiral properties correlated with the superconductivity, we investigated the field- and temperature-dependent $R_\textrm{xx}^{2\omega}$. Figure \ref{fig3}a and \ref{fig3}b show the $R_\textrm{xx}^{2\omega}-B$ curves in different temperatures under magnetic fields in the ab-plane and along the c-axis, respectively. The phase diagrams of $R_\textrm{xx}^{2\omega}$ with respect to the magnetic field and the temperature are shown in Fig. \ref{fig3}c and \ref{fig3}d, respectively. The dashed lines are the contours of the first harmonic resistance $R_\textrm{xx}^{\omega}$ at zero resistance 0$R_\textrm{N}$, half of the normal state resistance (0.5$R_\textrm{N}$), three-quarters of the normal state resistance(0.75$R_\textrm{N}$), and the onset of transition. The nonreciprocal magnetotransport mainly locates at the region between the contours of 0$R_\textrm{N}$ and 0.5$R_\textrm{N}$ when the field is in the $ab$-plane and along the $c$-axis. Therefore, the nonreciprocal response basically occurs at the vortex flow regions, indicating that the vortex motion could be the domination. Above the contour of $R=0.5R_\textrm{N}$, $R_\textrm{xx}^{2\omega}$ decays gradually because the magnetic field diminishes the pinning potentials and the vortices move more freely. Above the contour of $R=0.75R_\textrm{N}$, the free flow regions continuously evolved into the amplitude fluctuation region and nonreciprocity is almost indiscernible as demonstrated in Fig. \ref{fig3}a and \ref{fig3}b. However, the $R_\textrm{xx}^{2\omega}$ did not disappear until $R$ reaches 0.95$R_\textrm{N}$ (Fig. \ref{fig2}d) at 3.8 K, where the contribution of amplitude fluctuation of order parameter may not be neglected.

\begin{figure}[h]
\centering
\includegraphics[width=0.4\linewidth]{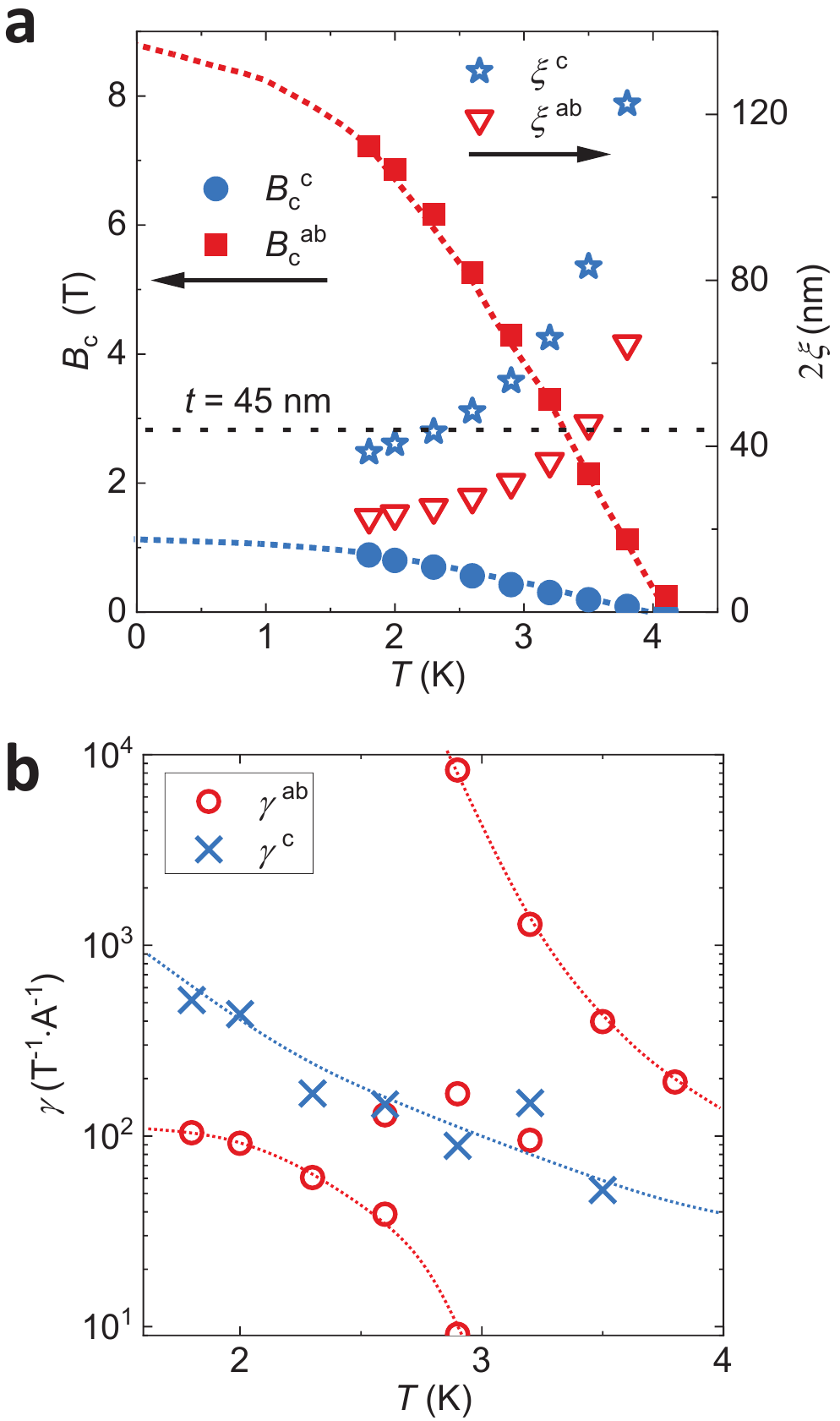}
\caption{
\label{fig4} 
\textbf{Coherence lengths and nonlinear coefficient.} \textbf{a} Critical field and coherence length are extracted for $B/\mkern-4mu/ab$-plane and $B/\mkern-4mu/c$-axis cases. Black dotted line shows the thickness of the sample. \textbf{b} The nonlinear coefficient $\gamma=\frac{2V^{2\omega}}{V^{\omega}BI}$ is calculated from several series of peaks.
}
\end{figure}

On the other hand, the $R_\textrm{xx}^{2\omega}$ under in-plane field is larger than that along the $c$-axis, suggesting an anisotropic nonreciprocal magnetotransport. Since the crystal is in a quasi-two-dimensional structure, anisotropic superconducting properties should also affect the vortex motion. We then examined the coherence lengths to estimate the size of the vortex from the anisotropic upper critical field $B_\textrm{c}$ (at $R=0.99R_\textrm{N}$). According to the Ginzburg-Landau theory \cite{tinkham2004introduction}, the in-plane and out-of-plane coherence lengths are $\xi_\text{ab}^{2}=\frac{\Phi_\textrm{0}}{2\pi\mu_\textrm{0}B_\text{c}^{2}}$ and $\xi_\textrm{c}^{2}=\frac{\Phi_\textrm{0}}{2\pi\mu_\textrm{0}\sqrt{B_\textrm{c}^\textrm{c}B_\textrm{c}^\textrm{ab}}}$, which are given in Fig.\ref{fig4} a, where $\Phi_\textrm{0}$ is flux quanta. Thus, the anisotropic factor can be estimated as $\lambda=\xi_\textrm{c}/\xi_\textrm{ab}$ (=1.72 at 1.8 K), indicating a weak anisotropy. Meanwhile, the thickness of the sample is comparable to the diameter of a vortex of $2\xi_\textrm{ab}$. Once $B/\mkern-4mu/ab$-plane, $R_\textrm{xx}^{2\omega}$ is enhanced due to the size effect and/or possible anisotropic pinning potential \cite{PhysRevB.98.054510}. Consequently, we conclude that the asymmetric property of pinning potential is three-dimensional.

Moreover, several specific series of peaks are observed other than one main peak predicted by the theory of vortex dynamics \cite{PhysRevB.98.054510}. Particularly, four series can be identified when $B/\mkern-4mu/ab$-plane and much more series exist in $B/\mkern-4mu/c$-axis case. The locations of each series of peaks are weakly temperature-dependent. In other words, the nonreciprocity is strong when some number of vortices exist in the device, indicating that the many-body interaction of vortices may also play a role \cite{de2006controlled}. We select three major series of peaks in Fig. \ref{fig3}a to estimate the $\gamma$ value as $\gamma=\frac{2V^{2\omega}}{V^{\omega}BI}$, and plot temperature dependence of $\gamma$ in Fig. \ref{fig4}b. While the magnitude of these three peaks is comparable in the order of several m$\Omega$ at 2.8 K, the coefficient $\gamma$ varied from about 5 T\textsuperscript{-1}A\textsuperscript{-1} to near 8000 T\textsuperscript{-1}A\textsuperscript{-1} for high-$H$ peaks and low-$H$ peaks, respectively. It is reasonable that the magnetic field weakens the pinning potential. The temperature plays a similar role as all these coefficients drastically decrease with increasing temperature. To compare with other systems, the values of $\frac{2V^{2\omega}}{V^{\omega}j}$, where $j$ is the current density, are also estimated as up to about $10^{-11}$m$^{2}$A$^{-1}$. We then conclude that the ratchet effect is remarkable in CsV\textsubscript{3}Sb\textsubscript{5}, comparable to the noncentrosymmetric 2D materials such as MoS\textsubscript{2} and NbSe\textsubscript{2}.

\subsection*{Current dependent nonreciprocity}

\begin{figure*}[h]
\centering
\includegraphics[width=0.8\textwidth]{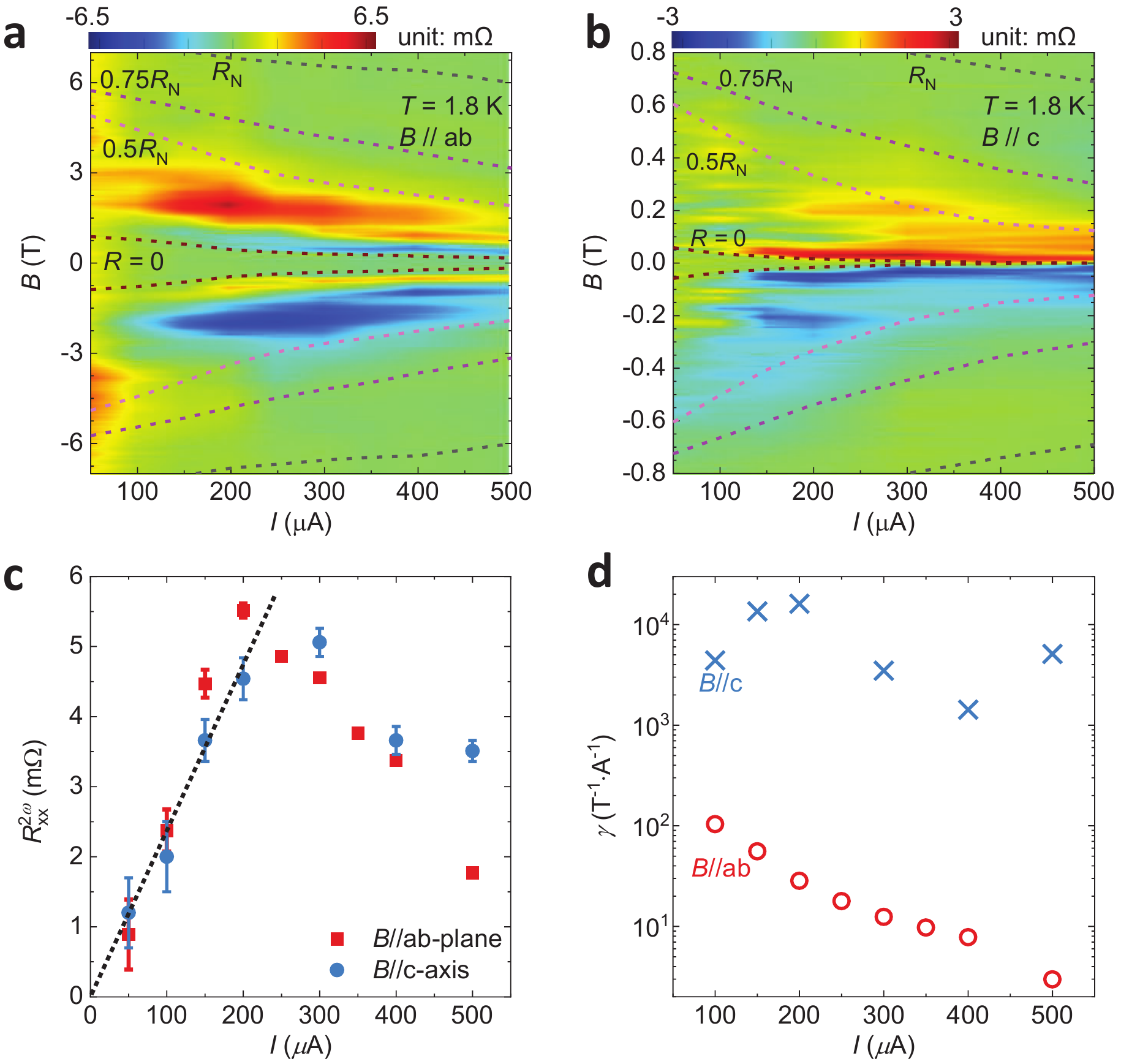}
\caption{
\label{fig5} 
\textbf{Current dependence of nonreciprocity of the 45-nm-thick sample.} \textbf{a} and \textbf{b} are color mapping of the $R_\textrm{xx}^{2\omega}$ as a function of magnetic field and current. The dashed lines are the contours of first harmonic resistance $R_\textrm{xx}^{\omega}$ of 0$R_\textrm{N}$ , 0.5$R_\textrm{N}$, 0.75$R_\textrm{N}$ and onset of transition. \textbf{c} Current dependence of the maximum second harmonic resistance for $B/\mkern-4mu/ab$-plane and $B/\mkern-4mu/c$-axis. Error bars are standard deviations of the resistances. \textbf{d} The corresponding coefficient $\gamma$ as estimated by $\gamma=\frac{2V^{2\omega}}{V^{\omega}BI}$.
}
\end{figure*}

We then focus on the current dependence of the nonreciprocity, which should follow the relation of $R_\textrm{xx}^{2\omega}\sim\gamma BI$. The phase diagrams of the first and the second harmonic signals are plotted in Fig. \ref{fig5}a and b, respectively. The effect of current on $R_\textrm{xx}^{2\omega}$ is similar to the effect of temperature as it suppresses the superconductivity and reduces the region of $R_\textrm{xx}^{2\omega}$. Meanwhile, several series of peaks and the corresponding sign reversals are observed as well. The maximum of the second harmonic resistance is extracted as a function of current as given in Fig. \ref{fig5}c, where the linear relation at current lower than 0.2 mA is verified for both $B/\mkern-4mu/ab$-plane and $B/\mkern-4mu/c$-axis cases. When applying a larger current, the signal starts to reduce because of the suppression of the rectification effect of vortex motion by the relative weakening of the pinning potentials \cite{MCAZhang,Itahashieaay9120,PhysRevB.72.064522,PhysRevB.42.2639,PhysRevMaterials.4.074003}. The current dependence of coefficient $\gamma$ of these maximum peaks behaves differently as shown in Fig. \ref{fig5}d, which drops exponentially as increasing the current when $B/\mkern-4mu/$ab-plane, while no such tendency can be observed clearly for $B/\mkern-4mu/$c-axis, possibly owing to the maximum peaks belonging to different series of peaks. It is worth noting that the nonreciprocity can be modified by current as well as $B$ and $T$, providing promising applications such as superconducting diodes \cite{MCAAndo}.

\subsection*{Discussion}

For the mechanism of the nonreciprocity phenomenon in superconductors, several explanations have been proposed for superconducting SrTiO\textsubscript{3}\cite{Itahashieaay9120}, PbTaSe\textsubscript{2}\cite{ideue2020giant}, Bi\textsubscript{2}Te\textsubscript{3}/FeTe interface\cite{MCAYasuda}, and NbN thin films\cite{nakamura2020}, which can be basically attributed to the intrinsic symmetry breaking through the vortex dynamics such as viscous vortex flow, ratchet effect, or specific mechanisms of superconductivity under spontaneous symmetry-breaking conditions. Under the vortex motion regime, one possibility is that the vortex liquid state and thermally activated vortex state may contribute to the response with opposite signs \cite{MCAZhang}. The viscous vortex flow is also possible to the nonreciprocal signal, which predicts a monotonic enhancement of $\gamma$ as decreasing temperature. Nevertheless, one can only observe an individual nonreciprocity peak for each mechanism instead of multi-peaks in our present results. Another mechanism is the effect of periodical pinning potential when the vortices match the lattice constant or artificially structured pinning sites, resulting in the modulation of condensing energy\cite{shiomi2017oscillatory,nagao2006periodic,swiecicki2012strong,lyu2021superconducting}. After breaking the inversion symmetry of the pinning potential, a well-controlled many-body interaction can realize sign reversal in artificial structures \cite{de2006controlled}. For single crystals, however, the symmetry of pinning potential should reflect the intrinsic symmetry of the crystal as well studied on MoS\textsubscript{2} \cite{MoS2PRR} and PbTaSe\textsubscript{2} \cite{ideue2020giant}, where the anisotropy of nonreciprocity follows that of crystal. Therefore, the appearance of series of nonreciprocity peaks in our present results suggests an intrinsic symmetry breaking as well.

One possible spontaneous symmetry breaking can be from the electronic structure, which is mainly dominated by the nature of the crystal structure and quantum phases. Hoshino $et$ $al.$ proposed a possible mechanism \cite{PhysRevB.98.054510} in noncentrosymmetric crystal structures and interfaces such as Rashba superconductor, and topological surface state with superconducting proximity effect. For the intrinsic asymmetry like spin-orbital effects and time-reversal symmetry breaking, an asymmetric energy dispersion relation can be induced as $E\left(s,\textrm{\textbf{k}}\right)\neq E\left(s,-\textrm{\textbf{k}}\right)$ and $E\left(s,\textrm{\textbf{k}}\right)\neq E\left(-s,-\textrm{\textbf{k}}\right)$, which is the key to the nonreciprocity. In bulk CsV\textsubscript{3}Sb\textsubscript{5}, the Rashba spin-orbital interaction seems unlikely to exist, because the crystal structure is centrosymmetric. Alternatively, a recent study of Raman spectrum in CsV\textsubscript{3}Sb\textsubscript{5} thin films revealed an intrinsic instability of crystalline structure \cite{liu2022anomalous}, and STM measurement also shows an inhomogeneous Cs-surface \cite{RN987}. Such degradation may reduce the symmetry and modify its electronic structure \cite{zhang2022emergence}, providing a possibility to induce nonreciprocity. Nevertheless, these effects ought to be weak since our crystals is relatively thick, and particularly, the samples are single-crystalline and well protected by BN, the degradation of crystalline can be basically avoided. 

In addition, the charge orders can also modulate the rotational, translation, and time reversal symmetry. In CsV\textsubscript{3}Sb\textsubscript{5}, the nematicity \cite{RN987} reveals a C$_\textrm{2}$ rotational symmetry, and the unidirectional 4$\bf{a}_\textrm{0}$ CDW also breaks rotational symmetry \cite{li2022rotation}. The amplitudes of three  \textbf{q} vectors within the in-plane 2$a_\textrm{0}\times$2$a_\textrm{0}$ CDW are chiral, which can break the time-reversal symmetry as well \cite{chiral,wang2021electronic}. However, the CDW-induced lattice distortion CsV\textsubscript{3}Sb\textsubscript{5}  was found to be centrosymmetric \cite{PhysRevX.11.041030}, and no any evidence suggests that one of these charge orders can break the inversion symmetry up to now.

Another possible origin of the spontaneously symmetric breaking is the unconventional superconducting state. Among a variety of possible pairing states, superconducting states on topological bands should keep a nontrivial nature and the pair potential demonstrates anisotropic, for instance a spinless $p$-wave superconductor \cite{mackenzie}. Previously, the nonreciprocal transport has been comprehensively studied on the interfacial systems of Bi$_2$Te$_3$/FeTe \cite{MCAYasuda}, which were attributed to inversion symmetry breaking of the topological superconductivity in the interface. The effect of other unconventional states, for example, chiral superconductors \cite{zinkl2021symmetry}, still needs further theoretical analysis and experimental evidence. Although the superconductivity symmetry of $A$V\textsubscript{3}Sb\textsubscript{5} family is still an open question as discussed in the introduction part, the ARPES and first principles calculations \cite{PhysRevLett.125.247002,HU2021,RN986,PhysRevLett.127.177001} indicate the existence of topological bands and surface states, and a recent work reported an observation of two-fold symmetry on the $c$-axis resistivity under in-plane rotation of magnetic field \cite{RN985}. Therefore, our nonreciprocal transport phenomenon should be a promising probe to reflect the anisotropic nature of the unconventional superconductivity state in CsV\textsubscript{3}Sb\textsubscript{5}.


In conclusion, we have carefully investigated the nonreciprocity under the dissipative state of CsV\textsubscript{3}Sb\textsubscript{5} superconductors. With the reduction of thickness, we found that the superconductivity is firstly enhanced and then suppressed due to the competition with CDW. Strong nonreciprocal signals can be detected when superconductivity occurs, whose strength is comparable to the artificially structured or noncentrosymmetric superconductors. The second harmonic resistance can be observed as the magnetic field both in the $ab$-plane and along the $c$-axis, which can be basically attributed to the vortex ratchet motion. The nonreciprocal signal with respect to magnetic fields splits into several series of peaks that are asymmetric with magnetic fields. We suggest the intrinsic symmetry breaking in superconducting states may induce the magnetochirality in the CsV\textsubscript{3}Sb\textsubscript{5} superconductor.

\section*{Methods}

\subsection*{Crystal growth}

High-quality single crystal were synthesized by Sb flux method and described in previous work\cite{wangqi}.

\subsection*{Device fabrication}

The thin crystals of CsV\textsubscript{3}Sb\textsubscript{5} were prepared by mechanical exfoliation from bulk crystal and transferred on pre-patterned SiO\textsubscript{2}(300 nm)/Si substrate using polydimethylsiloxane (PDMS) stamps in a laboratory-made vacuum transfer system. The pre-patterned leads were fabricated by standard photolithography method using Photoresist (AZ5214) and subsequent deposition of Ti/Au (5 nm/20 nm) by magnetron sputtering. Some of the samples were then covered by a $h$-BN capping layer for further degradation. The thicknesses of the samples were confirmed by AFM after transport measurement.

\subsection*{Transport measurements}

The transport measurement was carried out in PPMS. The four-terminal DC and AC signal was measured by set of Keithley 2400 and 2182a and set of Keithley 6221 and OE1022 lock-in amplifier, respectively. The first and second harmonic resistance were defined as $R^{\omega}=V^{\omega}/I_\textrm{0}$ and $R^{2\omega}=V^{2\omega}/I_\textrm{0}$, where $I_\textrm{0}$ is the amplitude of the AC current applied and $V^{\omega}$ and $V^{2\omega}$ are the amplitude of first and second harmonic voltage. The current frequency was set to be 113 Hz to lower the noise and the phase of second harmonic signal was set as $\pi/2$.

\section*{Data Availability Statement}

The data supporting the findings of this study are available from the corresponding authors upon reasonable request.
\section*{Acknowledgement}

This research was supported in part by the National Natural Science Foundation of China (Grants No. 12004251, 12104302, 12104303, U1932217, and 11974246), the Natural Science Foundation of Shanghai (Grant No. 20ZR1436100, 19ZR1477300), the Science and Technology Commission of Shanghai Municipality, the Shanghai Sailing Program (Grant No. 21YF1429200), the start-up funding from ShanghaiTech University, Beijing National Laboratory for Condensed Matter Physics, the Interdisciplinary Program of Wuhan National High Magnetic Field Center (WHMFC202124), the Elemental Strategy Initiative conducted by the MEXT, Japan, Grant Number JPMXP0112101001, JSPS KAKENHI Grant Number JP20H00354 and A3 Foresight by JSPS, and National Key R\&D Program of China (Grant No. 2018YFA0704300).

\section*{Competing Interests}

We declare no competing interest.

\section*{Author Contributions}

Y.W. and Q.W. contributed equally to this work. Y.W. and J.L. designed the research. Q.W. and Y.Q. grew the single crystals of CsV\textsubscript{3}Sb\textsubscript{5}. K.W. and T.T. grew the single crystals of $h$-BN. Y.W. fabricated the samples and performed the electrical transport measurements. Y.Z. performed the atomic force microscopy experiment. Y.Z., P.D., J.H., Y.D., Y.L., B.T., C.Z., and H.Z. assisted the device fabrication. Y.W., X.Z., J.W., Q.Y., and J.L. contributed to analyze the data. Y.W. and J.L. wrote the paper.







\section*{Figure Legends}

\end{document}